\documentclass[draftcls, onecolumn, journal, 12pt]{IEEEtran}

\usepackage{cite}
\usepackage{graphicx}
\usepackage{epsfig}
\usepackage{latexsym}
\usepackage{amssymb,amsmath}
\usepackage[usenames]{color}

\makeatletter
\def\ifundefined{\@ifundefined}
\makeatother

\def\mathlette#1#2{{\mathchoice{\mbox{#1$\displaystyle #2$}}%
                               {\mbox{#1$\textstyle #2$}}%
                               {\mbox{#1$\scriptstyle #2$}}%
                               {\mbox{#1$\scriptscriptstyle #2$}}}}
\renewcommand{\Vec}[1]{\mathlette{\boldmath}{#1}}

\newcommand{\be}{\begin{equation}}
\newcommand{\ee}{\end{equation}}
\newcommand{\ba}{\begin{array}}
\newcommand{\ea}{\end{array}}
\newcommand{\bdm}{\begin{displaymath}}
\newcommand{\edm}{\end{displaymath}}
\newcommand{\bea}{\begin{eqnarray}}
\newcommand{\eea}{\end{eqnarray}}
\newcommand{\bean}{\begin{eqnarray*}}
\newcommand{\eean}{\end{eqnarray*}}
\newcommand{\me}{\mathrm{e}}

\newcommand{\mj}{\mathrm{j}}

\def\argmin{\mathop{\mathrm{argmin}}}

\def\diag{\mathrm{diag}}
\def\var{\mathrm{var}}

\def\dif{\mathrm{d}\,}
\def\E{\mathop{\mathbb{E}}}

\def\nTD{\ensuremath{T_\mathrm{D}}} 
\def\nTS{\ensuremath{T_\mathrm{S}}} 

\def\nRS{\ensuremath{R_\mathrm{S}}} 

\def\oH{\ensuremath{^\mathrm{H}}} 
\def\oT{\ensuremath{^\mathrm{T}}} 
\def\oF{\ensuremath{^{(\mathrm{F})}}} 

\def\nfC{\ensuremath{f_\mathrm{C}}}        

\def\nnuD{\ensuremath{\nu_\mathrm{D}}}     

\def\bias{\ensuremath{\mathrm{bias}}}      
\def\var{\ensuremath{\mathrm{var}}}        
\def\MSE{\ensuremath{\mathrm{MSE}}}        
\def\SNR{\ensuremath{\mathrm{SNR}}}        

\def\rank{\ensuremath{\mathrm{rank}}}      

\def\nSigmaU{\ensuremath{\Vec{\Sigma}_\mathrm{U}}}        

\def\nSigmaN{\ensuremath{\Vec{\Sigma}_\mathrm{N}}}        

\newcommand\inSet[2]{\ensuremath{\in\mathbb{#1}^{\,#2}}} 

\newcommand{\figpath}{Pictures/}

\IEEEoverridecommandlockouts

\title{Asymptotic Capacity Results for Non-Stationary Time-Variant Channels Using Subspace Projections}

\author{Thomas Zemen\authorrefmark{1} and Stefan M. Moser
\thanks{Thomas Zemen ({\bf{corresponding author}}) is with the Forschungszentrum Telekommunikation Wien (ftw.), Donau-City-Str.{} 1/3, 1220 Vienna, Austria (email:~thomas.zemen@ftw.at, phone: +43 1 5052830-81, fax: +43 1 5052830-99). Stefan M. Moser is with the Signal and Information Processing Laboratory, ETH Z\"urich. This work was carried out with funding from K\emph{plus} in the ftw. project I0 'Signal and Information Information Processing'.}}

\begin{document}

\bibliographystyle{IEEEtran}

\markboth{IEEE Transactions on Information Theory, submitted (May 10, 2005).} {Zemen and Moser: Asymptotic Capacity Results for Non-Stationary Time-Variant Channels Using Subspace Projections}

\maketitle

\begin{abstract}
In this paper we deal with a single-antenna discrete-time flat-fading channel. The fading process is assumed to be stationary for the duration of a single data block. From block to block the fading process is allowed to be non-stationary. The number of scatterers bounds the rank of the channels covariance matrix. The signal-to-noise ratio (SNR), the user velocity, and the data block-length define the \emph{usable} rank of the time-variant channel subspace. The usable channel subspace grows with the SNR. This growth in dimensionality must be taken into account for asymptotic capacity results in the high-SNR regime. Using results from the theory of time-concentrated and band-limited sequences we are able to define an SNR threshold below which the capacity grows logarithmically. Above this threshold the capacity grows double-logarithmically.
\end{abstract}


\section{Introduction}
In this paper we deal with a single-antenna discrete-time flat fading channel. The fading process is assumed to be stationary and Gaussian for the duration of a single data block of length $M$. From block to block the fading process is allowed to be non-stationary, i.e., we assume the fading to be independent across blocks. We analyze the validity of subspace based channel models for obtaining high signal-to-noise ratio (SNR) capacity results.

In wireless communication system the time-variation of the communication channel is caused by Doppler effects due to user movement and multipath propagation. The Doppler bandwidth of the channels time-variation is defined by the velocity of the user. The symbol rate is typically much higher than the maximum Doppler bandwidth.

In \cite{Liang04} the capacity of a non-coherent block-stationary fading model is analyzed and the pre-log (i.e., the ratio of capacity to $\log \SNR$ in the limit when the $\SNR$ tends to infinity $\lim_{\SNR\rightarrow\infty}\frac{C(\SNR)}{\log \SNR}$ ) is derived. The pre-log depends critically on the rank of the covariance matrix of one fading block. In order to estimate this rank for realistic scenarios Liang and Veeravalli use in [1] a subspace projection on Fourier basis functions \cite{Sayeed98}. Their derivation, however, lacks several important points:
\begin{itemize}
\item A projection of the time-variant channel $h[\cdot]$ on a subspace makes only sense in a context where $h[\cdot]$ needs to be estimated, e.g., from the observed channel output symbols. This means that the estimation becomes dependent on the $\SNR$.

\item Any estimation problem is based on a criterion that allows to specify the quality of a particular estimate. In \cite{Liang04} no such criterion is given. We propose to use a mean square error (MSE) criterion.

\item Once the problem of defining a subspace is understood as estimation problem with a given quality-criterion, it becomes obvious that the choice of the basis functions strongly influences the estimation error. In particular, the Fourier basis expansion, i.e., a truncated discrete Fourier transform (DFT), has the following major drawback in describing a time-variant channel: the rectangular window associated with the DFT introduces spectral leakage \cite[Sec. 5.4]{Proakis96}, i.e., the energy from low frequency Fourier coefficients leaks to the full frequency range. Furthermore, the truncation causes an effect similar to the Gibbs phenomenon \cite[Sec. 2.4.2]{Proakis96}. This leads to significant phase and amplitude errors at the beginning and at the end of a data block, and hence to a deteriorated MSE.

\item The choice of the dimension of the subspace has to be made in the context of the estimation problem. Hence, the optimum dimension will depend on the $\SNR$ by the trade-off between the square bias and the variance \cite{Zemen04, Scharf91}.
\end{itemize}

We suggest two subspace projections that give a significantly better MSE than the Fourier basis expansion of \cite{Liang04}:
\begin{itemize}
\item The Slepian basis expansion, as introduced in \cite{Zemen04} based on concepts from the theory of time-concentrated and band-limited sequences \cite{Slepian78}, offers a square-bias reduction of more than one magnitude compared to the Fourier basis expansion \cite{Zemen04}.

\item The Karhunen-Lo\`eve expansion provides the optimum basis functions in terms of a minimum MSE when the second order statistics are known.
\end{itemize}

\subsection*{Contribution:}
\begin{itemize}
\item We adapt the results from \cite{Liang04} taking into account the SNR-dependent subspace dimension.

\item We define a SNR threshold below which the capacity grows logarithmically. Above this threshold the capacity grows double-logarithmically. This result allows to bridge the gap between the results from \cite{Liang04} and \cite{Lapidoth05}.
\end{itemize}

\subsection*{The rest of the paper is organized as follows:}
The notation is presented in Section \ref{seNotation} and the signal model for flat-fading time-variant channels is introduced in Section \ref{se:FlatFadingSignalModel}. We provide the time-variant channel description based on physical wave propagation principles in Section \ref{se:PhysicalWave}. Subspace based channel models are reviewed in Section \ref{se:SubspaceChannelDescr} building the foundation for the capacity results in Section \ref{se:Capacity} which build on \cite{Liang04}. In Section \ref{se:Relevance} numerical results for the begin of the double-logarithmic capacity region are presented. We draw conclusions in Section \ref{se:Conclusions}.

\section{Notation}
\label{seNotation}
\begin{tabular}{ll}
$\Vec{a}$ & column vector with elements $a[i]$\\
$\Vec{A}$ & matrix with elements $[\Vec{A}]_{i,\ell}$\\
$\Vec{A}\oT$, $\Vec{A}\oH$ & transpose, conjugate transpose of $\Vec{A}$ \\
$\diag(\Vec{a})$ & diagonal matrix with entries $a[i]$ \\
$\Vec{I}_Q$ & $Q\times Q$ identity matrix \\
$a^*$, $\left|a\right|$ & complex conjugate, absolute value of $a$ \\
$\lfloor a \rfloor$ & largest integer, lower or equal than $a\in\mathbb{R}$\\
$\lceil a \rceil$ & smallest integer, greater or equal than $a\in\mathbb{R}$\\
$\left\|\Vec{a}\right\|$ & $\ell_2$-norm of vector $\Vec{a}$
\end{tabular}

\section{Signal Model for Flat-Fading Time-Variant Channels}
\label{se:FlatFadingSignalModel}

We consider the transmission of a symbol sequence $d[m]$ with symbol rate $\nRS=1/\nTS$ over a time-variant flat-fading channel. The symbol duration is much longer than the delay spread of the channel $\nTS\gg\nTD.$ Discrete time is denoted by $m$. The channel in equivalent baseband notation $h(t,\tau)$ incorporates the transmit filter, the physical channel and the matched receive filter.

Hence, we consider a discrete time model where the received sequences $y[m]$ is given by the multiplication of the symbol sequence $d[m]$ and the sampled time-variant channel $h[m]\triangleq h(m\nTS,0)$ plus additional circular symmetric complex white Gaussian noise $z[m]$
\be
\label{eq:FlatSignalModel}
y[m]=\sqrt{\SNR}h[m]d[m]+z[m]\,,\qquad m\in\{0,\ldots,M-1\}\,.
\ee
The transmission is block oriented with block length $M$. The time-variant channel is assumed to be independent from block to block. We normalize the system so that the channel inputs $d[m]$ have power constraint\footnote{We consider here for mathematical convenience a peak-power constraint instead of the more common average-power constraint. We believe that this assumption does not influence the results very much. See also the results and remarks in \cite[Sec.~XI]{Lapidoth05}.} $|d[m]|^2=1$, the channel coefficients $h[m]$ are circular symmetric complex Gaussian distributed with zero mean, $h[m]\sim\mathcal{CN}(0,1)$, and  $z[m]$ is a sequence of independent and identically distributed (i.i.d.) $\mathcal{CN}(0,1)$ random variables. The term $\SNR$ presents the signal-to-noise ratio.

\section{Physical Wave Propagation Channel Model}
\label{se:PhysicalWave}

We model the fading process using physical wave propagation principles \cite{Hofstetter04a}. The impinging wave fronts at the receive antenna are caused by $P$ scatterers. The individual paths sum up as
\be
\label{eq:DiscretePaths}
h[m]=\sum_{p=0}^{P-1}a_p\me^{\mj 2\pi f_p \nTS m}=\sum_{p=0}^{P-1}a_p\me^{\mj 2\pi \nu_p m}\,.
\ee
Here $f_p\in\mathbb{R}$ is the Doppler shift of path $p$. For easier notation we define the normalized Doppler frequency as $\nu_p=f_p\nTS$. The attenuation and phase shift of path $p$ is denoted by $a_p\in\mathbb{C}$. The Doppler shift of each individual path $p$ depends on the angle of arrival $\alpha_p$ in respect to the movement direction, the users velocity $v$, and the carrier frequency $\nfC$,
\be
\nu_p=\frac{v \nfC}{c_0}\nTS\cos\alpha_p\,.
\ee
The speed of light is denoted by $c_0$. The one sided normalized Doppler bandwidth is given by
\be
\nnuD=\frac{v \nfC}{c_0}\nTS\geq|\nu_p|\,.
\ee
It is assumed that $a_p$ and $\nu_p$ for $p\in\{0,\ldots,P-1\}$ are independent of each other and that both parameter sets are i.i.d..

We assume a time-variant block fading channel model. In this model the path parameters $a_p$ and $\nu_p$ are assumed to be constant for the duration of a single data block with length $M$. Their realization for the next data block is assumed to be independent. The fading process for the duration of a single data block is stationary, however from block to block the fading is non-stationary. This assumption is validated by measurement results from \cite{Viering03, Viering03a}.

We define the covariance matrix
\be
\Vec{\Sigma}_h=\E\{\Vec{h}\Vec{h}\oH\}
\ee
where the channel coefficients for a single block with length $M$ are collected in the vector
\be
\Vec{h}=[h[0], h[1], \ldots, h[M-1]]\oT\,.
\ee

Referring to (\ref{eq:DiscretePaths}) it is clear that the rank of $\Vec{\Sigma}_h$ is given by
\be
\label{eq:RankScatterer}
\rank(\Vec{\Sigma}_h)=\min(P,M)=Q
\ee
for $v>0$.

\section{Subspace Channel Description}
\label{se:SubspaceChannelDescr}
In this paper, we deal with non-stationary time-variant channels. For channel estimation at the receiver side, noisy observations $y[m]$ for $m\in\{0,\ldots,M-1\}$ are used. Thus, for channel estimation at the receiver side the effective covariance matrix
\be
\label{eq:EffectiveCov}
\Vec{\tilde{\Sigma}}_h=\Vec{\Sigma}_h+\frac{1}{\SNR}\Vec{I}_M
\ee
is essential, which takes into account the noise level at the receiver side.

We consider a subspace based channel description which expands the sequence $h[m]$ in terms of $D$ orthogonal basis function $u_i[m]$ for $i\in\{0,\ldots,D-1\}$
\be
\label{eq:SlepianBE}
h[m]\approx\tilde{h}[m]=\sum_{i=0}^{D-1}u_i[m]\gamma_i\,,
\ee
where $m\in\{0,\ldots,M-1\}$.

Due to the orthogonality of the basis functions we can estimate the basis expansion coefficients $\gamma_i$ according to
\be
\gamma_i=\sum_{m=0}^{M-1}u_i^*[m]d^*[m]y[m]\,.
\ee
Knowledge of the data symbols $d[m]$ can be achieved through an iterative estimation and detection scheme \cite{Zemen04b}.

The purpose of a subspace based channel model is to minimize the mean square error (MSE) by selecting appropriate basis functions $u_i[m]$ \emph{and} the correct subspace dimension $D$. The MSE of the basis expansion is defined as
\be
\label{eq:MSEModel}
\MSE_M=\frac{1}{M}\sum_{m=0}^{M-1}\E\left\{\left|h[m]-\tilde{h}[m]\right|^2\right\}\,.
\ee
It is shown in \cite{Zemen04,Niedzwiecki00} that $\MSE_M$ can be described as the sum of a square bias term and a variance term
\be
\label{eq:MSEM}
\MSE_M=\bias^2_M+\var_M\,.
\ee
These two terms show different behavior with respect to the $\SNR$ and the subspace dimension $D$. The square bias $\bias^2_M$ is independent of the $\SNR$ and gets smaller with increasing subspace dimension $D$. While the variance $\var_M$ increases with $D$ and with the noise variance $1/\SNR$. Thus, the subspace dimension $D$ defines the bias-variance tradeoff for a given $\SNR$ level \cite{Scharf91}.

In the following sections we will review different possibilities for the definition of the subspace spanned by the basis functions $u_i[m]$ for $i\in\{0,\ldots,D-1\}$ and $m\in\{0,\ldots,M-1\}$ and for selecting the subspace dimension $D$.

\subsection{Karhunen-Lo\`eve Expansion}
\label{subSe:KarhunenLoeve}
In general, if the second order statistic $\Vec{\Sigma}_h$ of the fading process is known, the Karhunen-Lo\`eve expansion \cite{Papoulis91a} provides the optimum basis functions in terms of minimum $\MSE_M$. The basis functions for the Karhunen-Lo\`eve subspace are defined by
\be
\Vec{\Sigma}_h\Vec{u}_i=\lambda'_i\Vec{u}_i
\ee
where $\Vec{u}_i\inSet{C}{M}$ has elements $u_i[m]$. The eigenvalues $\lambda'_i$ are sorted in descending order. The variance in (\ref{eq:MSEM}) is given by
\be
\var_M=\frac{D}{M}\frac{1}{\SNR}\,,
\ee
and the square bias can be expressed as
\be
\bias^2_M=\sum_{i=D}^{Q-1}\lambda'_i\,,
\ee
where $Q$ denotes the rank of the covariance matrix $\Vec{\Sigma}_h$.

The subspace dimension that minimizes $\MSE_M$ for a given $\SNR$ is found as \cite{Scharf91}
\be
\label{eq:SubSpaceDim}
D=\argmin_{D'\in\{1,\ldots,Q\}}\left(\sum_{i=D'}^{Q-1}\lambda'_i+\frac{D'}{M}\frac{1}{\SNR}\right) \,\in\{1,\ldots,Q\} \,.
\ee
Thus, the subspace dimension $D$ is defined by taking into account the $\SNR$. A similar approach is used in \cite{Lapidoth05, Lapidoth03} for stationary time-variant channels.

We define the notation
\be
\Vec{U}_{a:b}=[\Vec{u}_a,\ldots, \Vec{u}_b]
\ee
which collects the eigenvectors $\Vec{u}_a, \Vec{u}_{a+1},\ldots,\Vec{u}_b$. Similarly we define the diagonal matrix
\be
\Vec{\Lambda}'_{a:b}=\diag([\lambda'_a,\ldots,\lambda'_b])\,.
\ee
With this definition we can partition $\Vec{\tilde{\Sigma}}_h$ (\ref{eq:EffectiveCov}) as
\be
\label{eq:UseableNoisePartitioning}
\Vec{\tilde{\Sigma}}_h=\underbrace{\Vec{U}_{0:D-1}\Vec{\Lambda}_{0:D-1}\Vec{U}_{0:D-1}\oH}_{\nSigmaU(\SNR)} + \underbrace{\Vec{U}_{D:Q-1}\Vec{\Lambda}_{D:Q-1}\Vec{U}_{D:Q-1}\oH + \frac{1}{\SNR}\Vec{I}_M}_{\nSigmaN(\SNR)}\,.
\ee
The subspace based channel estimation actually models the covariance matrix of the \emph{useable} channel subspace $\nSigmaU(\SNR)$ only. The noise subspace $\nSigmaN(\SNR)$ is suppressed. Thus, matrix $\nSigmaU(\SNR)$ is also the one that will be relevant for the capacity calculations in Section \ref{se:Capacity}.

\subsection{Fourier Basis Expansion}
Without detailed knowledge of $\Vec{\Sigma}_{h}$ it is possible to define a subspace based on the normalized Doppler bandwidth $\nnuD$ only. In this case the Fourier basis expansion is a first reasonable choice. The Fourier basis functions are defined as
\be
\label{eq:Fourier}
u\oF_i[m]=\frac{1}{\sqrt{M}}\me^{\mj2\pi \frac{i-D\oF/2+1/2}{M}m}\,,
\ee
for $i\in\{0,\ldots,D\oF\}$ and $m\in\{0,\ldots,M-1\}$. The dimension is selected as
\be
D\oF=2\lfloor\nnuD M\rfloor+1\,.
\ee

The drawbacks of the Fourier basis expansion due to spectral leakage \cite[Sec. 5.4]{Proakis96} and the Gibbs phenomenon \cite[Sec. 2.4.2]{Proakis96} are well known \cite{Zemen03b,Schnitter04}. The mismatch between the Fourier basis functions $u\oF_i[m]$ (\ref{eq:Fourier}), which have discrete frequencies $\frac{i-D\oF/2+1/2}{M}$, and the physical wave propagation model (\ref{eq:DiscretePaths}), where the normalized Doppler frequencies are real valued, lead to an high square bias \cite{Zemen04,Zemen03b}.

In order to find a better suited basis expansion model Zemen and Mecklenbr\"auker \cite{Zemen04} introduce the Slepian basis expansion for time-variant channel modeling. It is shown in \cite{Zemen04} that the Slepian basis expansion offers a square bias reduction of more than one magnitude compared to the Fourier basis expansion. The most important properties of the Slepian basis expansion are reviewed in the next section.

\subsection{Slepian Basis Expansion}
\label{se:SlepianBE}

Slepian \cite{Slepian78} answered the question which sequence is bandlimited to the frequency range $[-\nnuD,\nnuD]$ and simultaneously most concentrated in a certain time interval of length $M$. The sequence $u[m]$ we are seeking shall have their maximum energy concentration in an interval with length $M$
\be
\label{eq:EnergyConcentration}
\lambda(\nnuD,M)=\frac{\sum\limits_{m=0}^{M-1}|u[m]|^2}{\sum\limits_{m=-\infty}^{\infty}|u[m]|^2}\,,
\ee
while being bandlimited to $\nnuD$, hence
\be
\label{eq:bandlimitedsequence}
u[m]=\int\limits_{-\nnuD}^{\nnuD}U(\nu)\me^{\mj2\pi m\nu} \dif \nu
\ee
where
\be
U(\nu)=\sum\limits_{m=-\infty}^{\infty}u[m]\me^{-\mj2\pi m\nu}\,.
\ee
We see that $0\leq \lambda(\nnuD,M) \leq 1$.

The solution of this constrained maximization problem are the discrete prolate spheroidal (DPS) sequences \cite{Slepian78}. The DPS sequences $u_i[m,\nnuD,M]$ are defined as the real-valued solution of
\be
\label{eq:DPSDefinition}
\sum_{\ell=0}^{M-1}\frac{\sin(2\pi\nnuD(\ell-m))}{\pi(\ell-m)}u_i[\ell,\nnuD,M]= \lambda_i(\nnuD,M)u_i[m,\nnuD,M]
\ee
for $i\in\{0,\ldots,M-1\}$ and $m\in\{-\infty,\ldots,\infty\}$ \cite{Slepian78}. For the remainder of the paper we drop the explicit dependence of $u_i[m]$ and $\lambda_i$ on $\nnuD$ and $M$.

The DPS sequences are doubly orthogonal on the infinite set $\{-\infty,\ldots,\infty\}=\mathbb{Z}$ and the finite set $\{0,\ldots,M-1\}$. The eigenvalues $\lambda_i$ are clustered near one for $i\leq\lceil 2 \nnuD M\rceil$ and decay rapidly for $i>\lceil 2 \nnuD M\rceil$. Therefore, the signal space dimension \cite[Sec. 3.3]{Slepian78} of time-limited snapshots of a bandlimited signal is \emph{approximately} given by \cite{Landau62}
\be
\label{eq:ApproxSignalSpaceDimension}
D'=\lceil 2 \nnuD M\rceil+1\,.
\ee
Albeit their rapid decay the eigenvalues stay bounded away from zero for finite $M$.

For our application we are interested in $u_i[m]$ for the time index set $m\in\{0,\ldots,M-1\}$ only. We introduce the term Slepian sequences for the index limited DPS sequences and define the vector $\Vec{u}_i\inSet{R}{M}$ with elements $u_i[m]$ for $m\in\{0,\ldots,M-1\}$.

The Slepian sequences $\Vec{u}_i$ are the eigenvectors of matrix $\Vec{C}$
\be
\Vec{C}\Vec{u}_i=\lambda_i\Vec{u}_i
\ee
and the energy concentration measured by $\lambda_i$ are the associated eigenvalues \cite{Slepian78}. Matrix $\Vec{C}$ has elements
\be
\label{eq:DPSMatrix}
[\Vec{C}]_{i, \ell}=\frac{\sin\left[ 2\pi(i-\ell)\nnuD\right]}{\pi (i-\ell)}\,,
\ee
where $i,\ell\in\{0,\ldots,M-1\}$.

We note that $\Vec{C}$ has full rank. The eigenvalue asymptotics for large $M$ and large $i>2\nnuD M$ are given by \cite{Slepian78}
\be
\label{eq:EigenvalueAsympt}
\lambda_i\approx\frac{1}{1+\me^{\pi b}}\,,
\ee
where $b$ is implicitly given through
\be
i=\lfloor2\nnuD M +(b/\pi)\log M\rfloor\,.
\ee

In the next section we apply the subspace concept in order to obtain capacity results for non-stationary time-variant channels.

\section{Asymptotic Capacity Results for High SNR}
\label{se:Capacity}

In \cite{Liang04} capacity results based on the rank of the channel covariance matrix $\Vec{\Sigma}_h$ are derived. The authors in \cite{Liang04} use a subspace based channel description based on the Fourier basis expansion. However, the SNR dependent partitioning of $\Vec{\tilde{\Sigma}}_h$ in a useable subspace $\nSigmaU$ (\ref{eq:UseableNoisePartitioning}) and a noise subspace $\nSigmaN$ was not taken into account.

\subsection{General Case}
From (\ref{eq:SubSpaceDim}) we see that in the high-SNR limit the optimum dimension corresponds to the rank $Q$:
\be
\label{eq:UsableDimension}
\lim_{\SNR\rightarrow\infty}D=Q\,.
\ee
Taking the partitioning (\ref{eq:UseableNoisePartitioning}) into account we can distinguish the following two general cases:
\begin{enumerate}
\item Number of scatterers $P$ equal or larger than the block length $M$,
\be
P\geq M\,.
\ee
From (\ref{eq:UsableDimension}) and (\ref{eq:RankScatterer}) we see that in the high-SNR limit
\be
\lim_{\SNR\rightarrow\infty}D=M\,.
\ee
Thus, the useable channel subspace $\nSigmaU$ becomes full rank. In this case the results in \cite{Liang04} provides an upper capacity bound as
\be
C(\SNR)\leq \log\log \SNR -\gamma - 1 - \log \lambda'_{M-1} + o(1)
\ee
where $\gamma$ denotes Euler's constant which is defined as
\be
\gamma=-\int_0^\infty\me^{-y}\log y \,\,\dif y \approx 0.5772\ldots
\ee
A similar capacity bound for the stationary case is found in \cite{Lapidoth05}.

\item Number of scatterers $P$ smaller than the block length $M$,
\be
P< M\,.
\ee
From (\ref{eq:UsableDimension}) and (\ref{eq:RankScatterer}) we see that in the high-SNR limit
\be
\lim_{\SNR\rightarrow\infty}D=P<M\,.
\ee
Thus, the useable channel subspace $\nSigmaU$ is not full rank and the capacity grows asymptotically as \cite{Liang04}
\be
\lim_{\SNR\rightarrow\infty}\frac{C(\SNR)}{\log\SNR}=\frac{M-P}{P}\,.
\ee
\end{enumerate}

\subsection{Flat Doppler Spectrum}
We can obtain more detailed results based on the knowledge of the eigenvalues of the useable covariance matrix. The theory of time-concentrated and band-limited sequences provides such knowledge for the specific matrix $\Vec{C}$ (\ref{eq:DPSMatrix}), see Section \ref{se:SlepianBE}. We assume that the time-variant channel has a flat Doppler spectrum in the interval $[-\nnuD, \nnuD]$. If such a fading process is observed for the duration of a data block with length $M$, the covariance matrix results in
\be
\label{eq:flatfadingBlock}
\Vec{\Sigma}_h=\E\{\Vec{h}\Vec{h}\oH\}=\frac{1}{2\nnuD}\Vec{C}\,.
\ee
The elements of matrix $\Vec{C}$ are defined in (\ref{eq:DPSMatrix}).

We assume $P\geq M$. From (\ref{eq:EigenvalueAsympt}) we know the smallest eigenvalue of $\Vec{C}$. Due to the relation (\ref{eq:flatfadingBlock}) we can express the smallest eigenvalue of $\Vec{\Sigma}_h$ as
\be
\label{eq:SmallestEigenvalue}
\lambda_{M-1}'=\frac{1}{2\nnuD}\lambda_{M-1}\approx\frac{1}{2\nnuD}\frac{1}{1+\me^{\pi^2\frac{M(1-2\nnuD)-1}{\log  M}}}\approx\frac{1}{2\nnuD}\me^{-\pi^2\frac{M(1-2\nnuD)-1}{\log  M}}\,.
\ee
Thus, we obtain a capacity upper bound as
\be
C(\SNR)\le\log\log\SNR-\gamma-1-\log\frac{1}{2\nnuD}+\pi^2\frac{M(1-2\nnuD)-1 } {\log M}+o(1)\,.
\ee

We can use (\ref{eq:SubSpaceDim}) in order to find the $\SNR$ threshold where the usable channel subspace becomes full rank $D=M$. This threshold is the $\SNR$ level where the MSE for the $M-1$ dimensional subspace
\be
\MSE_M(M-1)=\lambda_{M-1}+\frac{M-1}{M}\frac{1}{\SNR}
\ee
becomes equal to the MSE for the $M$ dimensional subspace
\be
\MSE_M(M)=\frac{1}{\SNR}\,.
\ee
Thus we can write
\be
\lambda_{M-1}+\frac{M-1}{M}\frac{1}{\SNR_{\mathrm{th}}}=\frac{1}{\SNR_{\mathrm{th}}}
\ee
which results in
\be
\label{eq:SNRthreshold}
\SNR_{\mathrm{th}}=\frac{1}{M}\frac{1}{\lambda_{M-1}}\,.
\ee

Inserting (\ref{eq:SmallestEigenvalue}) in (\ref{eq:SNRthreshold}) we obtain the SNR threshold

\be
\label{eq:Threshold}
\SNR_{\mathrm{th}}=\frac{1}{M}\frac{1}{\lambda'_{M-1}}=\frac{2\nnuD}{M}\left(1+\me^{\pi^2\frac{M(1-2\nnuD)-1}{\log  M}}\right)\approx\frac{2\nnuD}{M}\me^{\pi^2\frac{M(1-2\nnuD)-1}{\log  M}}\,,
\ee
below which the channel subspace is rank deficient. Thus, for $\SNR<\SNR_{\mathrm{th}}$ the capacity increases logarithmically. For $\SNR>\SNR_{\mathrm{th}}$ the useable channel subspace is full rank and the capacity increase becomes double-logarithmic. A similar result for stationary time-variant channels is given in \cite{Lapidoth05}.

For fast fading channels where $\nnuD\rightarrow1/2$ all eigenvalues become identical $\lambda_0=\lambda_1=\ldots=\lambda_{M-1}=1$ and
\be
\lim_{\nnuD\rightarrow\frac{1}{2}}\SNR_{\mathrm{th}}=\frac{1}{M}\,.
\ee

In general it will be desirable to operate a communication system below $\SNR_{\mathrm{th}}$. In the next section we will provide some numerical evaluations for wireless communication systems.

\section{Relevance For Practical Wireless Communication Systems}
\label{se:Relevance}

In \cite{Viering03a} the stationarity properties of wireless communication channels are analyzed by means of measurements. It is found that the fading process is stationary for a certain distance $d$ expressed in multiples of the wavelength $\lambda=c_0/\nfC$,
\be
d\leq\delta_{\mathrm{stat}}\lambda\,.
\ee
The stationarity distance $d$ is the path length for which the set of scatterers that generate the fading process does not change. After the user has traveled a distance longer than $d$ a new set of scatterers become active. For urban environments $\delta_{\mathrm{stat}}\approx100$ \cite{Viering03a}.

For a given symbol duration $\nTS$ we obtain a block length $M$ that depends on the actual user velocity
\be
\label{eq:BlockLength}
M(\nnuD)=\left\lfloor\frac{\delta_{\mathrm{stat}}\lambda}{v\nTS}\right\rfloor=\left\lfloor\frac{\delta_{\mathrm{stat}}}{\nnuD}\right\rfloor\,,
\ee
and thus also on the normalized Doppler bandwidth $\nnuD$.

\begin{figure}
\centering
\includegraphics[scale=0.5]{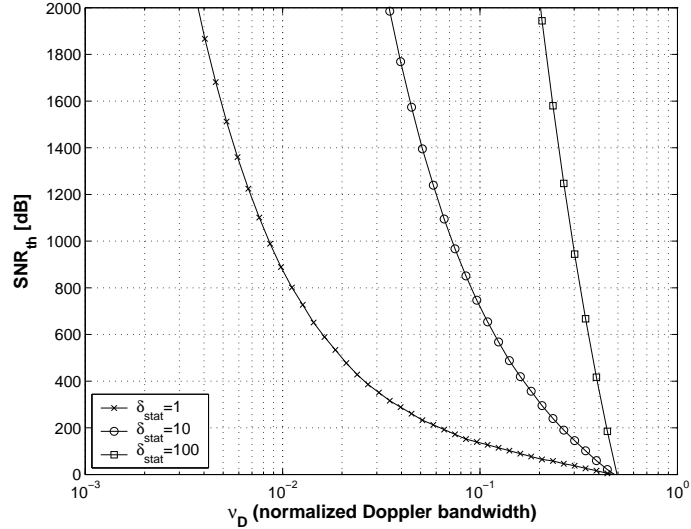}
\caption{Signal-to-noise ratio threshold $\SNR_{\mathrm{th}}$ that describes the boundary between logarithmic and double-logarithmic capacity growths. We plot results for a fading channel with stationarity time $\delta_{\mathrm{stat}}\in\{1,10,100\}$ (in multiples of wavelength). The normalized Doppler bandwidth $\nnuD$ is varied in the interval $(10^{-3}, 1/2)$.}
\label{SNRThreshold}
\end{figure}

Plugging (\ref{eq:BlockLength}) into (\ref{eq:Threshold}) we obtain
\be
\SNR_{\mathrm{th}}(\delta_{\mathrm{stat}},\nnuD)=\frac{2\nnuD}{\left\lfloor\frac{\delta_{\mathrm{stat}}}{\nnuD}\right\rfloor} \left(1+\me^{\pi^2\frac{\left\lfloor\frac{\delta_{\mathrm{stat}}}{\nnuD}\right\rfloor(1-2\nnuD)-1} {\log\left\lfloor\frac{\delta_{\mathrm{stat}}}{\nnuD}\right\rfloor}}\right)\,.
\ee
We evaluate $\SNR_{\mathrm{th}}(\delta_{\mathrm{stat}},\nnuD)$ for $\delta_{\mathrm{stat}}\in\{1,10,100\}$ and $\nnuD\in(0,1/2)$ in Fig. \ref{SNRThreshold}.

For current communication systems like UMTS at $f_C=2\,$GHz the normalized Doppler bandwidth is $\nnuD<0.02$. Thus, the double-logarithmic regime will not be reached. However, at higher carrier frequencies, e.g. $f_C=60\,$GHz or for underwater communication systems these bounds become more important.

\section{Conclusions}
\label{se:Conclusions}
We have shown that the number of scatterers $P$ bounds the rank of the channel covariance matrix. The $\SNR$, the user's velocity $v$, and the data block length $M$ define the \emph{usable} rank of the channel subspace.

The usable channel subspace grows with the SNR. This growth in dimensionality must be taken into account for asymptotic capacity results in the high-SNR regime. For $P<M$ the asymptotic capacity grows logarithmically, while for $P\geq M$ a double-logarithmic growth is obtained.

Using results from the theory of time-concentrated and band-limited sequences, we are able to specify the SNR above which the double-logarithmic capacity region starts in the special case of a flat Doppler spectrum.

\section{Acknowledgement}
We would like to thank Joachim Wehinger and Christoph F. Mecklenbr\"auker for helpful discussions.


\end{document}